\documentclass[aps,prb,twocolumn,superscriptaddress]{revtex4}
\usepackage{amssymb}
\usepackage{amsmath}
\usepackage{graphicx}
\usepackage{amsfonts}
\usepackage{bbold}
\usepackage{color}
\usepackage{epstopdf}

\begin{document}

\title{Transport properties of a two-lead Luttinger liquid junction out of
equilibrium: fermionic representation}
\author{D.N. Aristov}
\affiliation{Department of Physics, St.Petersburg State University, Ulianovskaya 1,
St.Petersburg 198504, Russia}
\affiliation{NRC "Kurchatov Institute", Petersburg Nuclear 
Physics Institute, Gatchina 188300, Russia}
\affiliation{Institute for Nanotechnology, Karlsruhe Institute of Technology, 76021
Karlsruhe, Germany }
\author{P. W\"olfle}
\affiliation{Institute for Nanotechnology, Karlsruhe Institute of Technology, 76021
Karlsruhe, Germany }
\affiliation{Institute for Condensed Matter Theory, Karlsruhe Institute of Technology,
76128 Karlsruhe, Germany}
\date{\today}

\begin{abstract}
The electrical current through an arbitrary junction connecting quantum
wires of spinless interacting fermions is calculated in fermionic
representation. The wires are adiabatically attached to two reservoirs at 
chemical potentials differing by the applied voltage bias. The relevant scale-dependent 
contributions in perturbation theory in the interaction up to infinite order
are evaluated and summed up. The result allows to construct
renormalization group equations for the conductance as a function of voltage
(or temperature, wire length). There are two fixed points at which the conductance 
follows a power law in terms of  a scaling variable $\Lambda$, which equals the bias voltage $V$,
if $V$ is the largest energy scale compared to temperature $T$ and inverse wire length $L^{-1}$, and 
interpolates between these quantities in the crossover regimes.
\end{abstract}

\pacs{71.10.Pm, 72.10.-d, 85.35.Be
}
\maketitle

\section{Introduction \label{sec:Intro}}

In the past few years, exactly one-dimensional quantum wires have become
available for experimental investigation in the form of carbon nanotubes,
chains of metal atoms or weakly side-coupled molecular chains in solids. The
new data emerging from these experiments \cite{Prokudina2014,Mebrahtu2013,
Jezouin2013}, in particular in non-equilibrium
situations, require a more detailed and more general theoretical description
than presently available. Electron transport in nanowires has been studied
theoretically for more than two decades. In the first papers it was found
that electron-electron interaction affects even the conductance of a clean
wire \cite{Apel1982,Kane1992}. In the case of realistic boundary conditions,
namely adiabatically attaching ideal leads to the interacting quantum wire,
the two-point conductance of a clean wire is that of the leads, equal to one
conductance quantum per channel, irrespective of the (forward scattering)
interaction \cite{Maslov1995}. The work of Kane and Fisher \cite{Kane1992} and Furusaki
and Nagaosa \cite{Furusaki1993} showed, that interaction has a dramatic
effect on the conductance in the presence of a potential barrier. Namely,
for repulsive interaction these authors found that the conductance tends to
zero as the temperature $T$, or more generally, the excitation energy of the
electrons approaches zero, while for attractive interaction the conductance
scales to its maximum value. This behavior has been shown to carry over to
the dependence on bias voltage, at sufficiently low temperatures (and for
long wires). 
There exists a large body of theoretical work addressing different aspects of transport through 
Luttinger liquids without or with impurities, such as the effect of the leads on a finite length wire, \cite{Safi1995,Furusaki1996} the response to an a.c.\ electric field \cite{Sassetti1996}, the appearance of oscillatory behavior in the nonlinear conductance \cite{Dolcini2003, Dolcini2005} and the emergence of bistability for the very strong interaction and bias voltage \cite{Egger2000}. The transport through Luttinger liquid junctions at not too strong interaction has also been calculated using the Functional Renormalization Group method as reviewed in   \cite{Metzner2012}. 
The results mentioned above have been mostly obtained within the
bosonization method, which needs to be amended by a correction taking care
of the physical situation of a wire of finite length attached to reservoirs
(see above). Experimentally, the predictions of theory have been found 
to be observed, at least qualitatively \cite{Bockrath1997,Milliken1996,Yacoby1996,
Tarucha1995,Tans1997}.

A proper treatment of the two-point conductance in the limit of weak
interaction, taking into account the gradual build-up of the Friedel
oscillations around the barrier as the infrared cutoff is lowered has been
given by Yue, Glazman and Matveev \cite{Yue1994}. These authors used the
perturbative RG for fermions to derive the conductance for an arbitrary (but
short) potential barrier ("fermionic representation"). In this paper we
extend the approach of Yue et al. to transport under stationary
non-equilibrium conditions. Following our extensive work on transport
in the linear regime
through junctions of Luttinger liquids at arbitrary strength of interaction 
\cite{Aristov2008,Aristov2009,Aristov2010,Aristov2011,
Aristov2011a,Aristov2012,Aristov2012a,Aristov2013}  we derive in the
following RG-equations for the conductance at finite bias voltage and for
any interaction strength. We use the fact that the $\beta$- function of the
RG-equation for the conductance can be obtained in very good approximation
by summing a class of contributions in perturbation theory in all orders of
the interaction \cite{Aristov2009}. A comparison of our previous results on
the linear response conductances of two- \cite{Aristov2009} and three-lead
junctions \cite{Aristov2010,Aristov2011a,Aristov2013} with or without
additional symmetries, or an applied magnetic flux \cite%
{Aristov2012a,Aristov2013}, with the results of the bosonization method, of
conformal field theory methods, of Bethe ansatz, where available, are in full
agreement provided those results were well-founded. In a few cases where the
conformal field theory result was based on an additional assumption we found
disagreement, which we interpret as saying that the assumption was not
justified.

In this paper we consider the transport of spinless fermions, which begs the question of how our results may be applied to experiment. The spinless Luttinger liquid model has actually been used to describe transport through spin-polarized quantum wires, as considered in Ref.\ \cite{Mebrahtu2013}    A generalization of our theory to spinful fermions is in progress.

\section{ The model \label{sec:Model}}

We consider a system of spin-less fermions in one dimension, interacting in
the region $a<|x|<L$ (the "wire"), adiabatically connected to reservoirs at $%
|x|>L$. There is a barrier in the narrow regime $|x|<a$ , which scatters the
fermions as described by the $S$-matrix (up to overall phase factors in the
individual wires) 
\begin{equation}
S=%
\begin{pmatrix}
r & t \\ 
\widetilde{t} & r%
\end{pmatrix}%
=%
\begin{pmatrix}
\sin \theta  & i\cos \theta e^{-i\varphi } \\ 
i\cos \theta e^{i\varphi } & \sin \theta 
\end{pmatrix}%
\end{equation}%
We choose this parametrization in terms of the transmission and reflection
amplitudes $t,r$ , since it is readily generalizable to the case of
multi-wire junctions ($n$ wires, $n>2$ ). The above form of the $S$-matrix
is completely general. 

In the continuum limit, linearizing the spectrum at the Fermi energy and
adding forward scattering interaction of strength $g_{j}$ in wire $j$ , we
may write the Hamiltonian $\mathcal{H}$ in the representation of incoming
and outgoing waves as 
\begin{equation}\begin{aligned}
\mathcal{H} &=\int_{0}^{\infty
}dx\sum_{j=1}^{2}[H_{j}^{0}+H_{j}^{int}\Theta (a<x<L)]\,, \\
H_{j}^{0} &=v_{j}\psi _{j,in}^{\dagger }i\nabla \psi _{j,in}-v_{j}\psi
_{j,out}^{\dagger }i\nabla \psi _{j,out}\,, \\
H_{j}^{int} &=2\pi v_{j}g_{j}\psi _{j,in}^{\dagger }\psi _{j,in}\psi
_{j,out}^{\dagger }\psi _{j,out}\,.
\end{aligned}\end{equation}%
We are using the chiral representation, labeling electrons in lead $j$ by $%
(j,\eta )\equiv j_{\eta }$\ where $\eta =+1$ for outgoing and $\eta =-1$\
for incoming electrons and all position arguments $x$ are on the positive
semi-axis. The range of the interaction lies within the interval $(a,L)$,
where $a>0$ serves as an ultraviolet cutoff (at energy scale $v_{j}/a$) and
separates the domains of interaction and potential scattering on the
junction; non-interacting leads are attached adiabatically at large $x$
beyond $L$. In terms of the doublet of incoming fermions $\Psi _{-}=(\psi
_{1,-},\psi _{2,-})$ the outgoing fermion operators may be expressed with
the aid of the $S$-matrix as $\Psi _{+}(x)=S\cdot \Psi _{-}(x)$ . For later
use we define density operators $\widehat{\rho }_{j,\eta =-1}=\psi
_{j,-}^{\dagger }\psi _{j,-}=\Psi _{-}^{+}\rho _{j}\Psi _{-}$, and $\widehat{%
\rho }_{j,\eta =1}=\psi _{j,+}^{\dagger }\psi _{j,+}=\Psi _{-}^{+}\widetilde{%
\rho }_{j}\Psi _{-}$\thinspace , where $\widetilde{\rho }_{j}=S^{+}\cdot
\rho _{j}\cdot S$\thinspace . The $2\times 2$\ matrices are defined by $%
(\rho _{j})_{\alpha \beta }=\delta _{\alpha \beta }\delta _{\alpha j}$ and $(%
\widetilde{\rho }_{j})_{\alpha \beta }=S_{\alpha j}^{+}S_{j\beta }$.  

\section{Charge current of free fermions \label{sec:FreeCurrent}}

The net current flowing outward through the point $z$ in wire $j$ is
composed out of two chiral components, moving towards ($\eta =-1$) and from (%
$\eta =1$) the junction, 
\begin{equation}
J_{j}(z)=v_{j}\Big(\langle \widehat{\rho }_{j,+}(z)\rangle -\langle \widehat{%
\rho }_{j,-}(z)\rangle \Big)
\end{equation}%
where  $v_{j}$ is the group velocity of the fermions. We use units where
electrical charge $e=1$, also $ \hbar =1$ and Boltzmann's constant $k_{B}=1.
$

We work with the Green's functions in this chiral basis and in Keldysh
formulation (we denote matrices in Keldysh space by an underbar), 
\begin{equation}
\underline{G}=%
\begin{pmatrix}
G^{R} & G^{K} \\ 
0 & G^{A}%
\end{pmatrix}%
\end{equation}

Here retarded, advanced and Keldysh components of the Green's functions, in
matrix form  in the chiral basis are given by ($2\times 2$ \ matrices in the
chiral basis are denoted by a hat, $\widehat{G}_{\eta _{l}\eta
_{j}}(l,y|j,x)=G(l,\eta _{l},y|j,\eta _{j},x)$) 
\begin{equation}\begin{aligned}
\widehat{G}_{\omega }^{R}(l,y|j,x) &=-\frac{i}{\sqrt{v_{l}v_{j}}}\theta
(\tau )e^{i\omega \tau }%
\begin{bmatrix}
\delta _{lj} & 0 \\ 
S_{lj} & \delta _{lj}%
\end{bmatrix}
\label{eq:Green} \\
\widehat{G}_{\omega }^{A}(l,y|j,x) &=\frac{i}{\sqrt{v_{l}v_{j}}}\theta
(-\tau )e^{i\omega \tau }%
\begin{bmatrix}
\delta _{lj} & S_{jl}^{\ast } \\ 
0 & \delta _{lj}%
\end{bmatrix}
\\
\widehat{G}_{\omega }^{K}(l,y|j,x) &=-\frac{i}{\sqrt{v_{l}v_{j}}}e^{i\omega
\tau }%
\begin{bmatrix}
\delta _{lj}h_{l} & S_{jl}^{\ast }h_{l} \\ 
S_{lj}h_{j} & S_{jm}^{\ast }S_{lm}h_{m}%
\end{bmatrix}
\\
\tau  &= \eta_{l} y/v_{l}-  \eta_{j} x/v_{j}
\end{aligned}\end{equation}%
where $h_{j}(\omega )=\tanh [(\omega -\mu _{j})/2T]$ is the equilibrium
distribution function in the reservoir of lead $j$ , characterized by the
chemical potential $\mu _{j}$ . We shall assume the temperatures $T$ in the
leads to be equal.

\begin{figure}[tbh]
\includegraphics[width=.5\columnwidth]{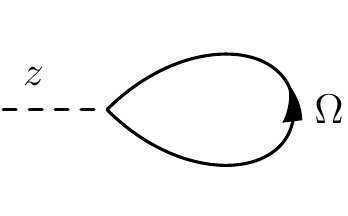}
\caption{ The diagram showing the zeroth order contribution to the current.
\label{fig:tadpole}}
\end{figure}

The average density of the chiral current at point $z$ , $\langle \rho
_{j,\eta }(z)\rangle $ , is represented by the diagram in 
Fig.\ \ref{fig:tadpole}. 

In terms of the Green's function matrix and defining the external vertex by
the Keldysh matrix $\underline{\gamma }_{ext}$

\begin{equation}
\underline{\gamma }_{ext}=\frac{i}{2}%
\begin{pmatrix}
1 & 1 \\ 
-1 & -1%
\end{pmatrix}%
\end{equation}%
\ we have

\begin{equation}
\langle \rho _{j,\eta }(z)\rangle =\int \frac{d\Omega }{2\pi } \mbox{Tr}_{K}\left[ 
\underline{\gamma }_{ext}\cdot \underline{G}_{\Omega }(j,\eta ,z|j,\eta ,z)%
\right]
\end{equation}%
with the trace $\mbox{Tr}_{K}$ taken over the Keldysh indices.

Using the expressions (\ref{eq:Green}), we obtain 
\begin{equation}\begin{aligned}
v_{j}\langle \rho _{j,-}(z)\rangle  &=\tfrac{1}{2}\int \frac{d\Omega }{2\pi 
}\,(1-h_{j}(\Omega )), \\
v_{j}\langle \rho _{j,+}(z)\rangle  &=\tfrac{1}{2}\int \frac{d\Omega }{2\pi 
}\,(1-\sum_{m}|S_{jm}|^{2}h_{m}(\Omega ))
\end{aligned}\end{equation}%
Notice here that the incoming current in the $j$th wire is characterized by
the distribution function referring to the same wire. The outgoing current
in the $j$th wire is characterized by the distribution functions referring
to the remaining wires. The dependence on $z$ vanishes in the d.c.\ limit
considered here.

Using the unitarity property (i.e.\ charge conservation), $%
\sum_{m}|S_{jm}|^{2}=1$, we may represent the net current in the form 
\begin{equation}
J_{j}^{(0)}(z)=\tfrac{1}{2}\int \frac{d\Omega }{2\pi }\,%
\sum_{m}|S_{jm}|^{2}(h_{j}(\Omega )-h_{m}(\Omega ))
\end{equation}%
which is a well-known expression. For the above choice of the weight
function $h_{j}(\Omega )$\ , the remaining integration can be easily done
with the result 
\begin{equation}
J_{j}^{(0)}(z)=\frac{1}{2\pi }\sum_{m}|S_{jm}|^{2}(\mu _{m}-\mu _{j})
\end{equation}%
The conductance\ (in units of the conductance quantum $e^{2}/2\pi \hbar $ )
of a two-lead junction is in lowest order given by 
\begin{equation}
G_{0}=J/V=|S_{12}|^{2}=t^{2}
\end{equation}%
where $V=\mu _{1}-\mu _{2}$ is the applied bias voltage. In the following we
will find it convenient to introduce the quantity $Y=1-2G_{0}$ characterizing
the conductance. 

\section{Current to first order in the interaction
\label{sec:1stOrder}}

\begin{figure}[htb]
\includegraphics*[ width=.8\columnwidth]{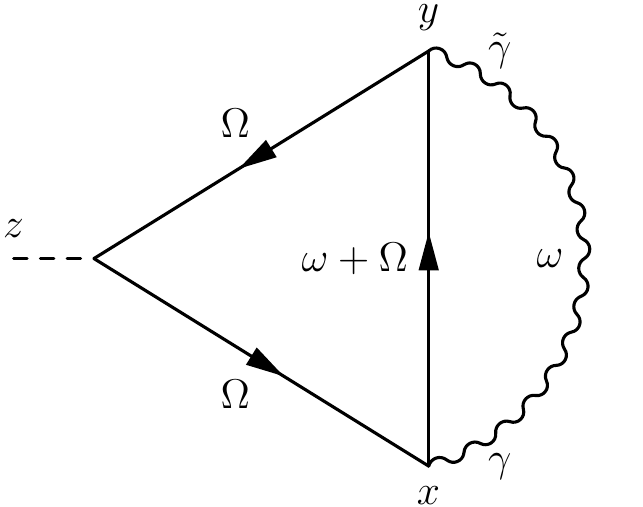}
\caption{ The diagram providing first-order correction to the current due to
interaction.}
\label{fig:triangle-diag}
\end{figure}

The first order correction to the current in the non-equilibrium case is
represented as the diagram depicted in Fig.\ \ref{fig:triangle-diag}. Here
the wavy line stands for the electronic interaction, taking place at the
point $x$ in the wire $l$. The contribution to the current of chirality 
$\eta _{n}$ in the $n$-th wire can be expressed as

\begin{equation}\begin{aligned}
J_{j_{\eta }}^{(1)}(z) &=v_{j}\int \frac{d\Omega \,d\omega }{(2\pi )^{2}}%
\,\int dx\,dy\sum_{\mu =1,2}\sum_{l,\eta _{l}}\,\mbox{Tr}_{K}
[\underline{\gamma }_{ext}     \\
& \times \underline{G}_{\Omega }(j_{\eta },z|l_{\eta },y)%
\underline{\bar{\gamma}}^{\mu }\underline{G}_{\Omega +\omega }(l_{\eta
},y|l_{\eta },x) \underline{\gamma }^{\mu }   \\
& \times \underline{G}_{\Omega }(l_{\eta },x|j_{\eta
},z)](2\pi ig_{l}v_{l})\delta (x-y)
\end{aligned}\end{equation}

The trace $\mbox{Tr}_{K}$ is over the lower (fermionic) Keldysh indices; the
fermion-boson vertices, $\gamma _{ij}^{\mu }\rightarrow \underline{\gamma }%
^{\mu }$, $\bar{\gamma}_{ij}^{\mu }\rightarrow \underline{\bar{\gamma}}^{\mu
}$, tensors of rank $3$\ defined in Keldysh space, are given by

\begin{equation}
\gamma^{1}_{ij} =\bar\gamma^{2}_{ij} = \tfrac1{\sqrt{2}} \delta_{ij}, \quad
\gamma^{2}_{ij} =\bar\gamma^{1}_{ij} = \tfrac1{\sqrt{2}} \tau^{1}_{ij},
\label{def:gamma}
\end{equation}
with $\tau^{1}$ the first Pauli matrix.

Notice that, similarly to the case of zeroth order in the interaction, the
factor $v_{j}$ at the external point $z$ is compensated by the prefactor
coming from the Green's function, Eq.\ (\ref{eq:Green}). If the point of the
observation $z$ lies outside the interacting region, $z>L$, then the
dependence on $z$ disappears in the outgoing current, $%
J_{j,+}^{(1)}(z>L)=J_{j}^{(1)}$, whereas the corrections to the incoming
current are alltogether absent, $J_{j,-}^{(1)}(z>L)=0$. In what follows we
discuss the corrections to the outgoing current. In view of the later
generalization involving an infinite summation of higher order terms it is
useful to represent the above first order expression as

\begin{equation}\begin{aligned}
J_{j}^{(1)} & =  i \int \frac{d\omega }{2\pi }\,\int dx\,dy\,\sum_{l_{\eta
},m_{\eta }} 
\\  & \times
\mbox{Tr}_{K}[\ \underline{T}_{\omega }(m_{\eta },y|l_{\eta },x;j,+,z)%
\underline{L}_{0,\omega }(l_{\eta },x|m_{\eta },y)]  \label{1stL0}
\end{aligned}\end{equation}%
where we recall the definitions 
\begin{equation*}
l_{\eta }=(l,\eta _{l})\,
\end{equation*}%
etc.  Here we defined a "boson propagator" representing the interaction line

\begin{equation}
\underline{L}_{0,\omega }(l,\eta _{l},x|m,\eta _{m},y)=(2\pi
g_{l}v_{l})\delta (x-y)\delta _{lm}\tau _{\eta _{l},\eta _{m}}^{1}%
\begin{pmatrix}
1 & 0 \\ 
0 & 1%
\end{pmatrix}%
\end{equation}%
and the quantity $T$ representing the triangle of Green's functions in Fig.\ %
\ref{fig:triangle-diag}

\begin{equation}\begin{aligned}
\label{def:T}
T_{\omega }^{ \nu\mu }&(
m_{\eta },y|l_{\eta },x ;j_{\eta},z) 
\\
&=v_{j}\int 
\frac{d\Omega }{2\pi }\mbox{Tr}_{K}\left[ \underline{\gamma }_{ext}\underline{%
\widehat{G}}_{\Omega }(j_{\eta},z|m_{\eta },y ) \underline{\bar{%
\gamma}}^{\nu } \right.  \\
&\times 
\left. \underline{\widehat{G}}_{\Omega +\omega }(m_{\eta },y|l_{\eta },x)%
\underline{\gamma }^{\mu }\underline{\widehat{G}}_{\Omega }(l_{\eta },x |j_{\eta},z) \right]\,,
\end{aligned}\end{equation}
this diagram should be combined with the one, where the arrows on the fermonic lines are reverted.

The triangle diagram is characterized by two Keldysh indices and thus is
subdivided into four blocks. Symbolically, we write 
\begin{equation*}
\mbox{Tr}_{K}[\ TL]=T^{11}L^{R}+T^{22}L^{A}
\end{equation*}%
anticipating that $T^{21}=0,L^{21}=0$ (to be shown later)\ .

When integrating over $\Omega $ in (\ref{def:T}) we find two generic
integrals. One of them is 
\begin{equation*}
\int d\Omega \,(h_{j}(\Omega +\omega )-h_{j}(\Omega ))=2\omega
\end{equation*}%
and the other is 
\begin{equation}
\int d\Omega \,[1-h_{j}(\Omega +\omega )h_{m}(\Omega )]=2F(\omega +\mu
_{m}-\mu _{j}).  \label{def:F}
\end{equation}%
For the above form of $h_{j}(\Omega )$, we have $F(x)=x\coth (x/2T)$.

As mentioned above there are no corrections to the incoming currents. In
addition to this observation we should recall Kirchhoff's law, stating the
conservation of charge. Given that the total incoming current is equal to
the total outgoing current, we should have $J_{1}+J_{2}=0$, which is indeed
confirmed by direct calculation.

Taking these facts into account, only the difference of the currents, 
$J^{(1)} = \frac12 (J_{2}^{(1)}-J_{1}^{(1)})$,  is of
interest. This involves the difference of the components of $T$ belonging to
different leads. Accordingly, for the case of two leads, we define the
weighted difference (denoted by the same symbol, $T$, but dependent on fewer
variables),


\begin{equation}\begin{aligned}
\label{Tdiff}
 T_{\omega }^{\mu \nu }(m_{\eta },y|l_{\eta },x) 
& = \frac{1}{2}[T_{\omega }^{\mu \nu }(m_{\eta },y|l_{\eta },x;1,+,z
>L)  \\ &  -T_{\omega }^{\mu \nu }(m_{\eta },y|l_{\eta },x;2,+,z>L)]
\end{aligned}\end{equation}

The  $4\times 4$ matrices appearing here are now direct products of $2\times 2$
matrices in chiral space (outer block structure) and $2\times 2$ matrices in
lead space (inner block structure, see Table \ref{tab:Indices}\ ) 
\begin{equation}\begin{aligned}
\label{oddT}
T^{11} _{}&=
  \frac{ r^{2}t^{2}}  {8\pi }     
 [F(\omega +V)-F(\omega -V)]%
\\ &\times
\Phi_{\omega}  (y)  \begin{bmatrix}
0 & 0 & 0 & 0 \\ 
0 & 0 & 0 & 0 \\ 
1 & -1 & 0 & 0 \\ 
-1 & 1 & 0 & 0%
\end{bmatrix}%
\Phi_{\omega}^{\ast}(x) 
\,, \\
T^{22} _{}&=-(T^{11}_{})^{\dagger } \vert_{x\leftrightarrow y}
\,,\quad T^{21}_{}=0\,.
\\ 
\Phi_{\omega} (x) &= 
diag\left [\frac{e^{-i\omega x/v_1}}{v_1} , \frac{e^{-i\omega x/v_2}}{v_2}  ,
\frac{ e^{i\omega x/v_1}}{v_1} , \frac{e^{i\omega x/v_2}}{v_2}   \right] 
\end{aligned}\end{equation}%
The vanishing of $T^{21}_{}$ implies that the Keldysh component of the
renormalized interaction, $L^{K}$, does not enter. Inserting the components of 
$T^{\mu \nu }_{}$ and $L_{0}$ into the expression (\ref{1stL0}) for the current 
for two equal wires, $g_{j}=g$, $v_{j}=v$, we find

\begin{equation*}
J^{(1)}=-\frac{ gt^{2}r^{2}}{\pi}\int_{0}^{\omega _{c}}\frac{d\omega }{\omega }%
[F(\omega +V)-F(\omega -V)]\sin ^{2}\frac{\omega L}{v}
\end{equation*}%
Here we apply an upper cut-off $\omega _{c}$ given in the microscopic model
either as $\omega _{c}=v/a$ as mentioned above or $\omega _{c}=W$, the band
width. The conductance as a function of voltage $V$, temperature $T$, wire
length $L$, is found from there as

\begin{equation}
G^{(1)}=-2g\, G_{0}(1-G_{0})\;\Lambda (V,T,L) \,.
\end{equation}%
Here we introduced the scaling variable $\Lambda $

\begin{equation}
\Lambda (V,T,L)=\int_{0}^{\omega _{c}}\frac{d\omega }{\omega }%
\frac{F(\omega +V)-F(\omega -V)}{V}\sin ^{2}\frac{\omega L}{v} \,.
\label{Lambda-gen}
\end{equation}%
The factor $\sin ^{2}(\omega L/v)]$ guarantees convergence of the integral
at $\omega <1/t_{0}=\pi v/L$. At $\omega >1/t_{0}$ we may average this
rapidly oscillating function and approximate $\sin ^{2}(\omega L/v)]\simeq
1/2$. Employing this and analogous procedures for the cases of small $V,L^{-1}$
or small $V,T$ we may approximate $\Lambda $ as

\begin{equation}
\Lambda (V,T,L)\simeq \ln \left( \frac{\omega _{c}}{\max \{V,T,v/L\}}\right)  \,.
\end{equation}

\begin{table}[tbp]
\caption{ Convention for the indices }
\label{tab:Indices}%
\begin{ruledtabular}
\begin{tabular}{c|cccc} 
$\alpha$ = & 1& 2 & 3 & 4  \\   \hline 
before/after &\multicolumn{2}{c}{B}&\multicolumn{2}{c}{A}   \\  
 wire \#  & 1 & 2  & 1 & 2 
\end{tabular}
\end{ruledtabular}
\end{table}

\section{Scale-dependent part of the current: Summation to infinite order in
the interaction \label{sec:summation}}

\begin{figure}[htb]
\includegraphics*[ width=.95\columnwidth]{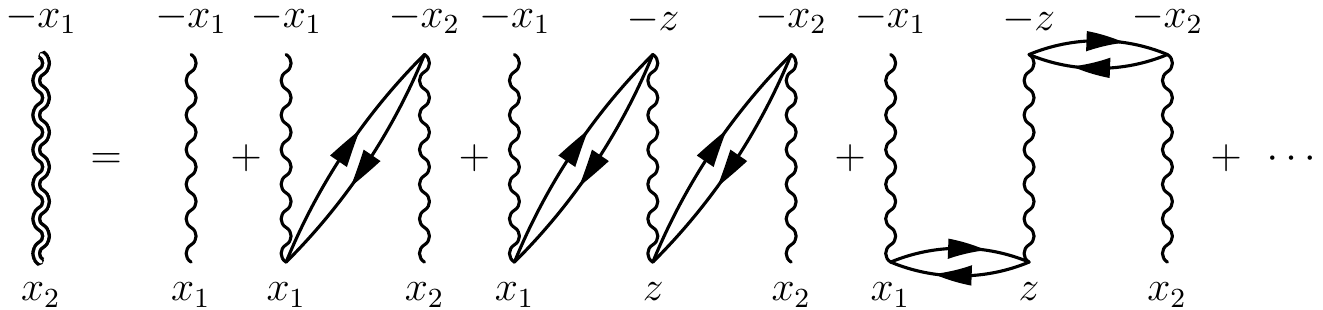}
\caption{ A series of diagrams showing the screening. The negative sign of the
coordinate corresponds to the incoming (B) electrons.}
\label{fig:ladder}
\end{figure}

As shown in our previous work, the perturbative treatment may be extended
into the strong coupling regime by summing up an infinite series of relevant
scale-dependent contributions to the conductance in all orders (``ladder approximation''). These
represent a self-energy renormalization of the ``boson propagator'' $L_{0}$
introduced above. They thus technically constitute a renormalized one-loop
contribution to the RG equation. This series can still be represented by the
generic diagram of Fig.\ \ref{fig:triangle-diag}, but the wavy line of
electronic interaction should be dressed by screening effects, as discussed
below.

As a result of this summation the interaction line, $g_{l}$, acquires
non-locality and retardation effects. Moreover, if we have initially only
diagonal components in Keldysh space, after the summation we generate a
Keldysh component and in general a rather complicated structure.
Schematically, we replace $L_{0}$ by $L$ in Eq.\ (\ref{1stL0}):

\begin{equation}
\underline{L}_{0}\rightarrow \underline{L}=%
\begin{pmatrix}
L_{\omega }^{R}(l_{\eta },x|m_{\eta },y) & L_{\omega }^{K}(l_{\eta },x|m_{\eta },y) \\ 
0 & L_{\omega }^{A}(l_{\eta },x|m_{\eta },y)%
\end{pmatrix}%
\end{equation}%
We now embark on the calculation of $\underline{L}$\ . Introducing compact
notation, we express the lowest order result $\underline{L}_{0}$ in the form 
$L_{0}^{\mu\nu}(l_{\eta },x|m_{\eta },y)=
\delta _{\mu\nu}\tau _{\eta _{l},\eta _{m}}^{1}\delta
_{lm}\, g\delta (x-y)=\mathbf{1}_{K}\otimes {\tau }_{C}^{1}\otimes \mathbf{1}%
_{w}\, g\delta (x-y)$. 


The integral equation describing the summation of the relevant infinite
class of diagrams (see Fig.\  \ref{fig:ladder}) takes the form

\begin{equation}\begin{aligned}
\underline{L} &=\underline{L}_{0}+\underline{L}_{0}\ast \underline{\Pi }%
\ast \underline{L}_{0}+\underline{L}_{0}\ast \underline{\Pi }\ast \underline{%
L}_{0}\ast \underline{\Pi }\ast \underline{L}_{0}+\ldots    \\
&=\underline{L}_{0}+\underline{L}_{0}\ast \underline{\Pi }\ast \underline{L}
\end{aligned}\end{equation}%
where $\underline{\Pi }$ represents a fermion bubble

\begin{equation}\begin{aligned}
\Pi _{\omega }^{ \mu \nu}(l_{\eta },x | j_{\eta },y) &= i \int \frac{d\Omega }{2\pi 
}\mbox{Tr}_{K}[\underline{\bar{\gamma}}^{\mu }\underline{G}_{\Omega +\omega
}(l_{\eta },x|j_{\eta },y)
\\ & \times
\underline{\gamma }^{\nu }\underline{G}_{\Omega
}(j_{\eta },y|l_{\eta },x) ]
\end{aligned}\end{equation}%
The multiplication $\ast $ is defined as

\begin{equation*}\begin{aligned}
(\Pi \ast L)_{\omega }^{\mu \nu }(j_{\eta },y|n_{\eta },x)&=\int
dz\sum_{l_{\eta }}\sum_{\lambda =1,2}\Pi _{\omega }^{\mu \lambda }(j_{\eta
},y|l_{\eta },z)
\\ & \times
L_{\omega }^{\lambda \nu }(l_{\eta },z|n_{\eta },x)
\end{aligned}\end{equation*}%
At the level of Keldysh structure we have

\begin{equation*}\begin{aligned}
\begin{pmatrix}
L^{R} & L^{K} \\ 
0 & L^{A}%
\end{pmatrix}
&=%
\begin{pmatrix}
L_{0} & 0 \\ 
0 & L_{0}%
\end{pmatrix}%
+ \\
&+%
\begin{pmatrix}
L_{0} & 0 \\ 
0 & L_{0}%
\end{pmatrix}%
\ast 
\begin{pmatrix}
\Pi ^{R} & \Pi ^{K} \\ 
0 & \Pi ^{A}%
\end{pmatrix}%
\ast 
\begin{pmatrix}
L^{R} & L^{K} \\ 
0 & L^{A}%
\end{pmatrix}%
\end{aligned}\end{equation*}%
which means that we can solve the integral equation in three steps.

First, we solve the coupled integral equations in the retarded sector 
\begin{equation}
L^{R}=L_{0}+L_{0}\ast \Pi ^{R}\ast L^{R}  \label{eq:LR}
\end{equation}%
Second, considering that 
\begin{equation*}
L^{A}=L_{0}+L_{0}\ast \Pi ^{A}\ast L^{A}
\end{equation*}
if we are using the relation between $\Pi ^{A}$ and $\Pi ^{R}$, we need not
solve this equation separately. Third, we notice for completeness that 
\begin{equation*}
L^{K}=L_{0}\ast \Pi ^{R}\ast L^{K}+L_{0}\ast \Pi ^{K}\ast L^{A}
\end{equation*}%
and hence, 
\begin{equation*}\begin{aligned}
L^{K} &=(1-L_{0}\ast \Pi ^{R})^{-1}\ast L_{0}\ast \Pi ^{K}\ast L^{A} \\
&=L^{R}\ast \Pi ^{K}\ast L^{A}
\end{aligned}\end{equation*}%
where we used (\ref{eq:LR}) to obtain the second equality. This means that
once we have $L^{R}$, we can easily find the two remaining components, $L^{A}
$ and $L^{K}$. We recall, however, that as pointed out above the component $%
L^{K}$ does not enter the calculation of the current.

The solution for $L^{R}$ in the linear response case was presented
previously in our work \cite{Aristov2012}. We follow that derivation but
reformulate it somewhat for the present purposes. First we define the
integral (scalar) kernel 
\begin{equation}\begin{aligned}
P _{\omega }(j,x|l,z) &=(2\pi v_{j}v_{l})^{-1}(\delta (\tau )+i\omega \theta
(\tau )e^{i\omega \tau }) \\
\tau  &=x/v_{j}-z/v_{l}
\end{aligned}\end{equation}%
and the matrix quantity 
\begin{equation}\begin{aligned}
\label{PiR}
\widehat{\mathbf{\Pi} }^{R} &= 
\begin{pmatrix}
  \mathbf{\Pi } (-x|-z)  , & \mathbf{Y}^{T} \Pi (-x|z) \\ 
\mathbf{Y} \Pi (x|-z) , &   \mathbf{\Pi } (x|z)  %
\end{pmatrix}%
\end{aligned}\end{equation}
where   $ \Pi _{jl}(x|z) = \delta_{jl}P _{\omega }(j,x|l,z) $, $\mathbf{Y} \Pi (x|z)  = Y_{jl} P _{\omega }(j,x|l,z)$ and 
$\mathbf{Y}^{T} \Pi (x|z)  = Y_{lj} P _{\omega }(j,x|l,z)$ with  $Y_{jl}=|S_{jl}|^{2}$. 
In the case of two leads we have $\mathbf{Y}=\mathbf{Y}^{T}$.
Notice that $\mathbf{Y}^{T} \Pi (-x|z) = 0 $ for $x,z>0$, and we use the full form  (\ref{PiR}) for future reference.

Then we express the integral equation for $\mathbf{L}^{R}$ as a 
$2\times 2$ matrix equation in the chiral space
\begin{equation}\begin{aligned}
&\widehat{\mathbf{L}}^{R}(x|y)=2\pi \delta (x-y)%
\begin{pmatrix}
0 & \mathbf{g} \\ 
\mathbf{g} & 0%
\end{pmatrix}
  \\
&-2\pi \int_{a}^{L}dz%
\begin{pmatrix}
\mathbf{g}\mathbf{Y} \Pi (x|-z) & \mathbf{g} \mathbf{\Pi} (x|z) \\ 
\mathbf{g} \mathbf{\Pi}  (-x|-z) & 0%
\end{pmatrix}%
\widehat{\mathbf{L}}^{R}(z|y)\,,
\end{aligned}\end{equation}%
Here $\widehat{\mathbf{L}}^{R}$ is a $4\times 4$ (in the general case of $n$
leads $2n\times 2n$ ) matrix.  The elements of the $2\times 2$ matrices in
chiral space (denoted by a hat) are $2\times 2$ matrices in the space of the 
$2$ leads (denoted by bold letters). The matrix of interaction constants is
then given by 
$\mathbf{g}=diag[ g_{1}v_{1} , g_{2}v_{2}  ] $. 
The scattering properties of the junction are
encoded in the $2\times 2$ matrix $\mathbf{Y}$. 
 The equation for $\mathbf{L}^{A}$ is similar to the  above, but 
\begin{equation*}
\widehat{\mathbf{\Pi} }^{A}  =
  \big( \widehat{\mathbf{\Pi} }^{R} \big) ^{\dagger} 
\Big|_{x \leftrightarrow z  } = 
\begin{pmatrix}
0 & \mathbf{1} \\ 
\mathbf{1} & 0%
\end{pmatrix}%
\widehat{\mathbf{\Pi} }^{R}%
\begin{pmatrix}
0 & \mathbf{1} \\ 
\mathbf{1} & 0%
\end{pmatrix}%
\Big|_{\omega \rightarrow -\omega ,\ \mathbf{Y}\rightarrow \mathbf{Y}^{T}}
 \end{equation*}%
Because $L_{0}$ does not contain $\omega $, $Y$, it follows that 
\begin{equation}
\widehat{\mathbf{L}}^{A}=
  \big( \widehat{\mathbf{L} }^{R} \big) ^{\dagger} 
\Big|_{x \leftrightarrow y  }
\label{LAviaLR}
\end{equation}%

The Keldysh part of the kernel takes the form presented in the appendix \ref%
{KeldyshPi}. We show there, that $\Pi^{K}$ is an even function of $V$, 
and therefore $L^{K}$ does not contribute to the current. 

Following the method of solution of the integral equation described in \cite%
{Aristov2012} we first solve the equation for the case $Y_{jl}=0$ , to give
a partial summation resulting in a auxiliary quantity $\mathbf{C}$

\begin{equation}\begin{aligned}
&\widehat{\mathbf{C}}(x|y)=2\pi \delta (x-y)%
\begin{pmatrix}
0 & \mathbf{g} \\ 
\mathbf{g} & 0%
\end{pmatrix}
  \\
&-2\pi \int_{a}^{L}dz%
\begin{pmatrix}
0 & \mathbf{g}\mathbf{\Pi} (x|z) \\ 
\mathbf{g}\mathbf{\Pi} (-x|-z) & 0%
\end{pmatrix}%
\widehat{\mathbf{C}}(z|y)\,,
\label{eq:defC}
\end{aligned}\end{equation}%
In terms of $\widehat{\mathbf{C}}$ the integral equation for $\widehat{%
\mathbf{L}}^{R}$ may be expressed as 
\begin{equation}\begin{aligned}
\label{eq:LviaC}
\widehat{\mathbf{L}}^{R}(x|y) & =\widehat{\mathbf{C}}(x|y)  
-2\pi \int_{a}^{L}dz_{1}dz_{2}\widehat{\mathbf{C}}(x|z_{1})%
\\ & \times
\begin{pmatrix}
0 & 0 \\ 
\mathbf{Y} {\Pi} (z_{1}|-z_{2}) & 0%
\end{pmatrix}%
\widehat{\mathbf{L}}^{R}(z_{2}|y)\,,
\end{aligned}\end{equation}%
The solution of the integral equation for $\widehat{\mathbf{C}}$, which is
of the Wiener-Hopf type,  may be found by an appropriate ansatz described in 
\cite{Aristov2012}. 
By construction, $\widehat{\mathbf{C}}(x|y)$ is diagonal in wire space, 
$\widehat{\mathbf{C}}_{jl}(x|y) = \delta_{jl}\widehat{C}_{j}(x|y)$. 
The explicit expressions for $\widehat{C}_{j}(x|y)$ are presented in the
Appendix \ref{sec:FullCL}.

Returning to the integral equation (\ref{eq:LviaC})  for $\widehat{\mathbf{L}}^{R}$ in terms
of $\widehat{\mathbf{C}}$ we observe that its kernel is separable and thus
the solution may be readily obtained. The explicit expressions 
and some details of the derivation of this result are given in Appendix \ref%
{sec:FullCL}.

\begin{figure}[htb]
\includegraphics*[ width=.8\columnwidth]{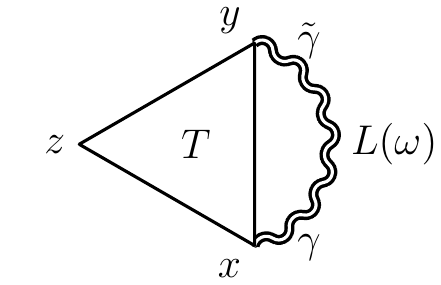}
\caption{ The schematic diagram, with already algebraic quantity $T(\protect%
\omega)$ and dressed interaction line $L(\protect\omega)$.}
\label{fig:triangle-diag2}
\end{figure}

An inspection of Eqs.\ (\ref{1stL0}), (\ref{oddT}) shows that the $x$, $y$ dependence of $T^{\mu\nu}_{}$ comes only from the matrices $\Phi_{\omega}^{\ast}(x)$,  $\Phi_{\omega}(y)$.  
We combine these matrices with $\widehat{\mathbf{L}}^{R}$ and integrate over the position
\begin{equation}
\widehat{\mathbf{L}}^{R}_{simple}= 
\int_{a}^{L} dx\,dy\, 
\Phi_{\omega}^{\ast} (x) \, 
\widehat{\mathbf{L}}^{R}(x|y)\, \Phi_{\omega}  (y) \,.
\end{equation}
Making now use of relations   (\ref{oddT}), (\ref{LAviaLR}), 
we arrive at a much simpler algebraic expression for 
the current. 
Instead of  (\ref{1stL0}) we have 
\begin{equation}
J^{(L)}  = -2\, \mbox{Im} \int \frac{d\omega }{2\pi }\,  \mbox{Tr }  [\widehat{T}_{core} ^{11}
\widehat{\mathbf{L}}^{R}_{simple} ] 
\label{TLmatrix}
\end{equation}%
with $T^{\mu\nu}_{core}$ obtained by putting $\Phi_{\omega} (x) =\Phi_{\omega} (y) = \mathbf{1}$  in Eq.\ (\ref{oddT}).
We show the algebraic relation  (\ref{TLmatrix}) diagrammatically in Fig.\ \ref{fig:triangle-diag2}.  

Introducing the quantities $d_{j}$ and $q_{j}$ 
\begin{equation}\begin{aligned}
\label{def:dq}
d_{j}^{2} &=1-g_{j}^{2}  \,, \\
q_{j}^{-1} &=\frac{g_{j}}{1+id_{j}\cot (\frac{\omega L}{v_{j}d_{j}})} \,,  %
\end{aligned}\end{equation}%
we  present
the simpler expressions of $\widehat{C}$, $\widehat{L}$\ integrated over
position.

 \begin{equation}\begin{aligned}
 \label{integL}
\widehat{ {L}}^{R}_{jk,simple} & = \frac{2\pi i }{\omega} \left(
  \delta_{jk} \widehat{ {C}} _{j,simple}  
 \right . \\ &  \left .
  +\Upsilon _{jk}   \begin{bmatrix}
  V_{1,j} V_{2,k} , & V_{1,j}V_{1,k} \\ 
V_{2,j}V_{2,k} , & V_{2,j}V_{1,k}%
  \end{bmatrix}
  \right)
\end{aligned}\end{equation}
where 
 \begin{equation}\begin{aligned} 
 \label{integCV}
 \widehat{ {C}} _{j,simple} &= 
\left(  \begin{bmatrix} -1, & 0 \\ 0,& -1 \end{bmatrix} 
\right . \\ & \left.
+ q_{j}^{-1} 
\begin{bmatrix} \frac{i d_{j}e^{-iL\omega/v}}{g_{j} \sin(\omega L/v_{j}d_{j})}, & 1 \\ e^{-2iL\omega/v},&
\frac{i d_{j}e^{-iL\omega/v}}{g_{j} \sin(\omega L/v_{j}d_{j})} \end{bmatrix}
\right ) \,, \\
 V_{1,j} &  =    ( \widehat{ {C}} _{j,simple} )_{12}  \,, \quad 
  V_{2,j} = ( \widehat{ {C}} _{j,simple} )_{11}  \,, \\
  \Upsilon _{jk} &=Y_{jl}(1-Q^{-1}\cdot Y)_{lk}^{-1},\quad Q_{jk}^{-1}=\delta
_{jk}q_{j}^{-1}, 
\end{aligned}\end{equation}

Combining the above results, (\ref{oddT}), (\ref{TLmatrix}), (\ref{integL}), (\ref{integCV}),  
we find the current for two equal wires, $g_{j}=g$, $v_{j}=v$, as

\begin{equation}\begin{aligned}
\label{ladderCurrent}
J^{(L)}=&\frac{g}{8\pi} \int_{0}^{\omega _{c}}d\omega \frac{F(\omega +V)-F(\omega -V)}{%
\omega }
\\ & \times
\mbox{Re}\left\{ \frac{2(1-Y^{2})}{1-gY+id\cot \left( \frac{L\omega 
}{dv}\right) }\right\} 
\end{aligned}\end{equation}%
Again the $\frac{1}{\omega }-$singularity of the integrand leads to a
logarithmically divergent contribution, which we identify as a scaling
contribution. The singularity is controlled by the largest of the three
energy scales, (i) energy scale $\omega _{L}=v/L$ controlled by the length $L
$, (ii) temperature $\omega _{T}=T$ , (iii) bias voltage $\omega _{V}=V$. In
the limit $V\rightarrow 0,T\rightarrow 0$ we have $F(\omega +V)-F(\omega
-V)=2Vsign(\omega )$. The $\frac{1}{\omega }-$singularity is in this case
cut off at the scale $\omega _{L}=dv/L$ \ by the $\cot \left( \frac{L\omega 
}{dv}\right) -$term in the denominator. Above this scale we may average the
rapidly oscillating function in the curly brackets in (\ref{ladderCurrent}) over one
oscillation period, $\omega _{0}<\omega <$ $\omega _{0}+(\pi /t_{0})$ ,  with $%
t_{0}=L/dv$ :

\begin{equation}
\frac{t_{0}}{\pi }\int_{\omega _{1}}^{\omega _{1}+\pi /t_{0}}\frac{d\omega }{%
1-gY\pm id\cot \omega t_{0}}=(1-gY+d)^{-1} \,,
\label{average}
\end{equation}%
such that the correction to the conductance is obtained as  
\begin{equation*}
G^{(L)}=- g\frac{(1-Y^{2})}{1-gY+d}\ln \frac{L}{a}
\end{equation*}%
in agreement with \cite{Aristov2009}. 
In the general case we find accordingly

\begin{equation}
G^{(L)}=- g\frac{(1-Y^{2})}{1-gY+d}\Lambda 
\label{Gladder}
\end{equation}%
where $\Lambda =\ln (\omega _{c}/\max \{V,T,v/L\}$ . In the limit of long wires, $L\to \infty$, a closed expression is found in Appendix \ref{sec:Lambda} in the form
\begin{equation}
\label{def:Lambda}
\Lambda =   \ln\left( \frac{\omega _{c} }{2\pi T} \right)  
  -  \mbox{Re}
\left[ \psi \left( 1 + \frac{iV}{2\pi T} \right)  
\right]    \,,
\end{equation} 
with   $\psi(x)$ digamma function. This function shows a smooth interpolation between the regimes 
with $\ln\left( \frac{\omega _{c} }{2\pi T} \right) + 0.577$ at $ V\ll T$ and 
$\ln\left( {\omega _{c} }/{V} \right) $ at $ V\gg T$.

Further corrections not considered here are generated by the Hartree diagrams of the self energy: in the nonequilibrium situation the local chemical potential is renormalized by a molecular field term involving the bare interaction and the local particle density. In Ref.\  \cite{Egger2000} this effect is included by applying a corresponding boundary condition to the thermodynamic Bethe ansatz fields. As a consequence a bistability of the current has been found at very strong interaction. We have not included this correction term into our analysis, as it would require a separate calculation of the single particle Green?s function, especially of the local chemical potential shift, which is beyond the scope of the present paper. We are therefore confining our considerations from weak up to moderately strong interaction, such that $K > 0.2$.

\section{Renormalization group equation for the conductance
\label{sec:deriveRG}}

The above calculation of the leading scale-dependent contribution to the
current allows us to derive a renormalization group (RG) equation for the
conductance\ $G=I/V$ as a function of the scaling variable $\Lambda =\ln
(\omega _{c}/\max [V,T,v/L])$ , $G=G(\Lambda )$. We thereby use the scaling
property of $G$ , $G(V,T,v/L,G_{0};g)=G(\Lambda ;g)$\ .  In our previous
works \cite{Aristov2009,Aristov2011} we explicitly checked this property in
the equilibrium situation. We directly calculated all the contributions to
the conductance up to third order in the interaction, which involves about $%
10^{4}$ diagrams.  It was shown that the principal contribution near the
fixed points (FPs) of this equation is obtained in one-loop order, with the
interaction being dressed as described above, $g\rightarrow L$. The scaling
exponents obtained this way are identical to those found earlier by the
method of bosonization.

Away from the FPs one finds in general additional non-universal
contributions, appearing first in the third order. These determine the
prefactor in the scaling law near the FP and also fine details of the
conductance in the intermediate regime. In the present study focused on the
transport far out of equilibrium it would be too costly to perform a similar
direct computation of all contributions up to third order. Instead we assume
that even out of equilibrium we have the scaling property and the scaling
exponents are fully determined by the contribution provided by the
approximation of fully dressing the interaction line of the one-loop
calculation.

This assumption may be justified by at least two facts.  First, in the renormalized one-loop calculation presented here the bias voltage $V$ appears as an infrared cutoff in the scale dependent terms, replacing the cutoff  energy scales temperature $T$ or level splitting $v/L$ present in equilibrium. It may therefore be expected that the structure of the scale-dependent terms generated by the cutoff $V$ is analogous to that of the terms generated by the cutoff $T$ or $v/L$.  No additional scale dependent terms are found in non-equilibrium and none of the scale-dependent terms present in equilibrium disappears in non-equilibrium. This suggests that the structure of the scale-dependent terms is preserved and therefore the scaling property is preserved even out of equilibrium. Secondly, as will be shown below, the results of our theory are in agreement with exact results obtained by other  methods.

We now briefly review the logic by which the RG-equation is derived from the
perturbative result. We start from the result for the renormalized
conductance $G$ as a power series expansion in the interaction, and
dependent on the scattering properties of the junction (encapsulated in the
conductance $G_{0}$ in the absence of interaction) obtained above, which
takes the general form ,  

\begin{equation}
G=G_{0}-gf(g,G_{0})\Lambda +\mathcal{O(}g^{2}\Lambda ^{2})
\label{generalpert}
\end{equation}%
In the approximation of summing up the leading terms in each order,
considered above, a very good approximation $f_{app}$ of the function $f(g,G)
$ has been obtained, see Eq.(\ref{Gladder}). We do not calculate the terms of order 
$g^{2}\Lambda ^{2}$ and higher in this paper. The relation
 Eq.\ (\ref{generalpert}) is valid in the asymptotic regime $g\Lambda \rightarrow 0$ .
With the aid of the scaling property we may find the analytic continuation
to finite values of $g\Lambda $. To this end we first invert the relation
(making use of $G=G_{0}+\mathcal{O}(g\Lambda )$ ) and write 
\begin{equation*}
G_{0}(g,G;\Lambda )=G+gf(g,G)\Lambda +\mathcal{O}(g^{2}\Lambda ^{2})
\end{equation*}%
Formally $G_{0}$ here is a function of $G,g$ and $\Lambda $ .\ We now employ
the crucial property that the value of the bare conductance, $G_{0}$, should
not depend on the scaling variable $\Lambda $\ , which means 
\begin{equation}
0=\frac{\partial G_{0}}{\partial \Lambda }+\frac{\partial G_{0}}{\partial G}%
\frac{dG}{d\Lambda }
\end{equation}%
and hence 
\begin{equation}
\frac{dG}{d\Lambda } =-\frac{gf(g,G)+\mathcal{O}(g^{2}\Lambda )}{1+g \Lambda [\partial
f(g,G)/\partial G] + \mathcal{O}(g^{2}\Lambda^{2} ) } 
\label{betaLambda}
\end{equation}
The scaling property of $G$ implies that the explicit $\Lambda$-dependence in (\ref{betaLambda}) cancels. This leads to the definition of the RG $\beta $-function
\begin{equation}
\frac{dG}{d\Lambda } =\beta (g,G)=-gf(g,G) \,.
\end{equation}%

 Our earlier direct third order
calculation in \cite{Aristov2009,Aristov2011} showed that the above ratio
was indeed independent of $\Lambda $ to the considered accuracy $g^{3}$. The
function $gf(g,G)$ has been calculated beyond the ladder approximation in 
\cite{Aristov2009} for the present case of a two-lead junction with the
result 
\begin{equation}
\tfrac{d}{d\Lambda }G=-gf_{app}(g,G)+c_{3}g^{3}G^{2}(1-G)^{2}+\mathcal{O}%
(g^{4})\,  \label{RGeq-generic}
\end{equation}
The second term here of order $g^{3}$originates from terms not contained in
the perturbation series for $L$ considered above. This term is subleading in
the sense that it vanishes more rapidly on approach to the fixed points at $%
G=0,1$ than the first term and does therefore not influence the critical
properties. There are indications that this is also the case with the higher
order contributions not captured by the ladder summation. 

A similar conclusion regarding the relative unimportance of corrections
beyond the ladder summation $g\rightarrow L$ was reached in \cite%
{Aristov2011} for the more general case of the three-lead Y-junction. In the
symmetrical setup the Y-junction was characterized by two conductances, and
after extensive computer analysis of perturbative corrections we arrived at
a set of two coupled RG equations. We found that the three-loop corrections,
not contained in the ladder series of diagrams, did not contribute to the
scaling exponents.

We expect that non-universal contributions to the $\beta -$function will
also exist in the case of non-equilibrium, but those terms will again be
unimportant when it comes to determine the critical behavior at the fixed
points. We will therefore approximate the exact function $f(g,G)$ by the one
determined in the ladder approximation and given through eq.(\ref%
{generalpert}), which gives rise to the $\beta -$function

\begin{equation}
\frac{d}{d\Lambda }G= -4 g\frac{G(1-G)}{1-g(1-2G)+d}\,
\label{eq:beta1}
\end{equation}
Introducing the Luttinger parameter $K = \sqrt{(1-g)/(1+g)}$, Eq.\ (\ref{eq:beta1}) may be re-expressed as 
\begin{equation}
\frac{d}{d\Lambda }G =-2 (1-K)\frac{G(1-G)}{K +(1- K) G}\,,
\label{eq:beta2}
\end{equation}
which is explicitly solved in the next section. 

\section{solution of RG equation \label{sec:RGsolution}}
Inverting Eq.\  (\ref{eq:beta2}) we write 
\begin{equation}
2 (1-K) d\Lambda =- d G \left (  \frac{K}{G } +\frac{1}{1-G} 
\right )
\,   ,
\end{equation}
which is integrated with the result
\begin{equation}
  \frac {1-G}{G^{K}} =   \frac{1-G_{0}} {G_{0}^{K}} e^{2 (1-K) \Lambda} 
  \label{explicitGLambda}
\,   .
\end{equation}

It is more instructive to exclude here the bare conductance, $G_{0}$,  and to represent our result as (cf.\ \cite{Jezouin2013})
\begin{equation}
  \frac {(1-G)/G^{K} |_{V_{1},T_{1}}}{(1-G)/G^{K} |_{V_{2},T_{2}}} 
    =    \frac { e^{2 (1-K) \Lambda (V_{1},T_{1}) }}  { e^{ 2 (1-K) \Lambda (V_{2},T_{2}) } }
  \label{ratio-conduc}
\,   .
\end{equation}

The latter exponential can be written as 
\[ e^{2 (1-K) \Lambda} = \left(\frac{\omega _{c}}{\max \{V,T,v/L\}} \right) ^{2 (1-K)} \,.
\] 
We see that near the two fixed points of the RG equation, $G=0$, $G=1$, we have the well-established scaling behavior \cite{Kane1992}
\begin{equation}\begin{aligned}
G & \simeq  \left(V/  {\omega _{0}} \right) ^{2 (K^{-1}-1)} \,, \quad G \to0 \,, \\ 
1- G & \simeq c_{\ast}\,  \left( {\omega _{0}}/{V } \right) ^{2 (1- K )}  \,, \quad G \to1 \,.
\label{GviaV}
\end{aligned}\end{equation} 
with an appropriate $\omega_{0}$ and where $V$ should be replaced by $\exp(\Lambda(V,T,v_{F}/l))$ in the more general situation. At the same time,  (\ref{explicitGLambda}) provides a smooth crossover between the fixed points, i.e.\  for those values of $G$ which, strictly speaking, are inaccessible by the bosonization approach.  

We notice further, that if the overall energy scale $\omega_{0}$ is fixed near one fixed point,  then the constant $c_{\ast}$ is entirely defined by the  three-loop and higher-loop terms in the RG equation. 
In the approximation of neglecting the three-loop terms, as in Eqs.\ (\ref{eq:beta2}), (\ref{def:Lambda}), the coefficient $c_{\ast} = 1$.  Keeping the three-loop terms, Eq.\ (\ref{explicitGLambda}) is approximately given by \cite{Aristov2009}
\begin{equation}
  \frac{G^{K}} {1-G} \left(1+G\tfrac{1-K}{K}\right)^{4c_{3}(1-K)} =  
  \left[\frac{\max \{V,T,v/L\}} {\omega _{0}}\right] ^{2 (1-K)}  
  \label{explicitG-c3}
\,   ,
\end{equation} 
which implies $c_{\ast} = K ^{-4c_{3}(1-K)}$. 

\subsection{Comparison with the exact solution at $K=1/2$}
To understand better the limitations of our  formula (\ref{explicitGLambda}), we compare it with the exact result at $K=1/2$. 
Explicitly, our expression in this case reads as 
\begin{equation}
G= 1- \frac{\sqrt{1+4x^{2}}-1}{2x^{2}}, \quad 
x = \frac{T}{T^{\ast}} \exp{\mbox{Re }
\psi \left[ 1 + \tfrac{iV}{2\pi T} \right]
} \,. \label{ourG12}
\end{equation}

The exact formula, obtained with the aid of the
  Bethe ansatz \cite{Weiss1995} is 

\begin{equation}
G_{1/2} = 1- \frac{4\pi T^{\ast} }{V} \mbox{Im } \psi \left[ \frac12 + \frac{T^{\ast}}{T}+ i \frac {V}{4\pi T}\right]
\,,
\end{equation}
with $T^{\ast}$ depending on the impurity backscattering amplitude and the ultraviolet cutoff. 
In two important limiting cases we have for the linear conductance 
\begin{equation}\begin{aligned}
G_{1/2}(T) & = G_{1/2}(V\to0,T) \,, \\ 
  &= 1- \tfrac{  T^{\ast} }{T}   \psi' \left( \tfrac12 + \tfrac{T^{\ast}}{T} \right) \,, 
\label{G12V0T} \\ 
  & \simeq \frac{1}{12} \left(\frac T{  T^{\ast} }\right)^{2}, \quad T \ll T^{\ast} \,,\\
 & \simeq 1- \frac{\pi^{2}}2 \frac{  T^{\ast} }{T} , \quad T \gg T^{\ast} \,.
\end{aligned}\end{equation}
And for the nonlinear conductance
\begin{equation}\begin{aligned}
G_{1/2}(V) & =  G_{1/2}(V,T\to0) \,, \\
&= 1- \tfrac{ 4\pi T^{\ast} }{V}  \arctan \tfrac{V}{4\pi T^{\ast}} \,,
\label{G12VT0}  \\ 
  & \simeq \frac{1}{12} \left(\frac V{  2\pi T^{\ast} }\right)^{2}, \quad V \ll T^{\ast} \,, \\ 
 & \simeq 1- \pi \frac{ 2\pi T^{\ast} }{V} , \quad V \gg T^{\ast} \,.
\end{aligned}\end{equation}
These expressions indicate the existence of non-universal three-loop terms in the RG $\beta$-function. 
Indeed, fixing the overall scale at small $T$  by $G_{1/2}(T) = (T/T^{\ast}_{1})^{2}$ with $T_{1}^{\ast} = \sqrt{12} T^{\ast}$ gives the above constant $c_{\ast}=\pi^{2}/(4\sqrt{3}) \simeq 1.424 $. At the same time, fixing the scale at small $V$ by $G_{1/2}(V) = (V/T^{\ast}_{2})^{2}$ with $T_{2}^{\ast} = 2\pi \sqrt{12} T^{\ast}$ produces $c_{\ast}=\pi/(2\sqrt{3}) \simeq 0.91$. 

This means, firstly, that the three-loop term $\sim c_{3} g^{3} G^{2}(1-G)^{2}$  in the RG equation (\ref{RGeq-generic}) has a \emph{different} prefactor  $c_{3}$,  
depending on whether the choice of low-energy cutoff is $T$ or $V$.
This fact was noted in \cite{Aristov2009} on the basis of direct computation of perturbative corrections. From the above estimate  $c_{\ast} = K ^{-4c_{3}(1-K)}$ we retrieve $c_{3} \simeq 0.255$ and $c_{3}\simeq -0.070$  for $G_{1/2}(T)$ and $G_{1/2}(V)$,  respectively. 

\begin{figure}[htb]
\includegraphics*[ width=.8\columnwidth]{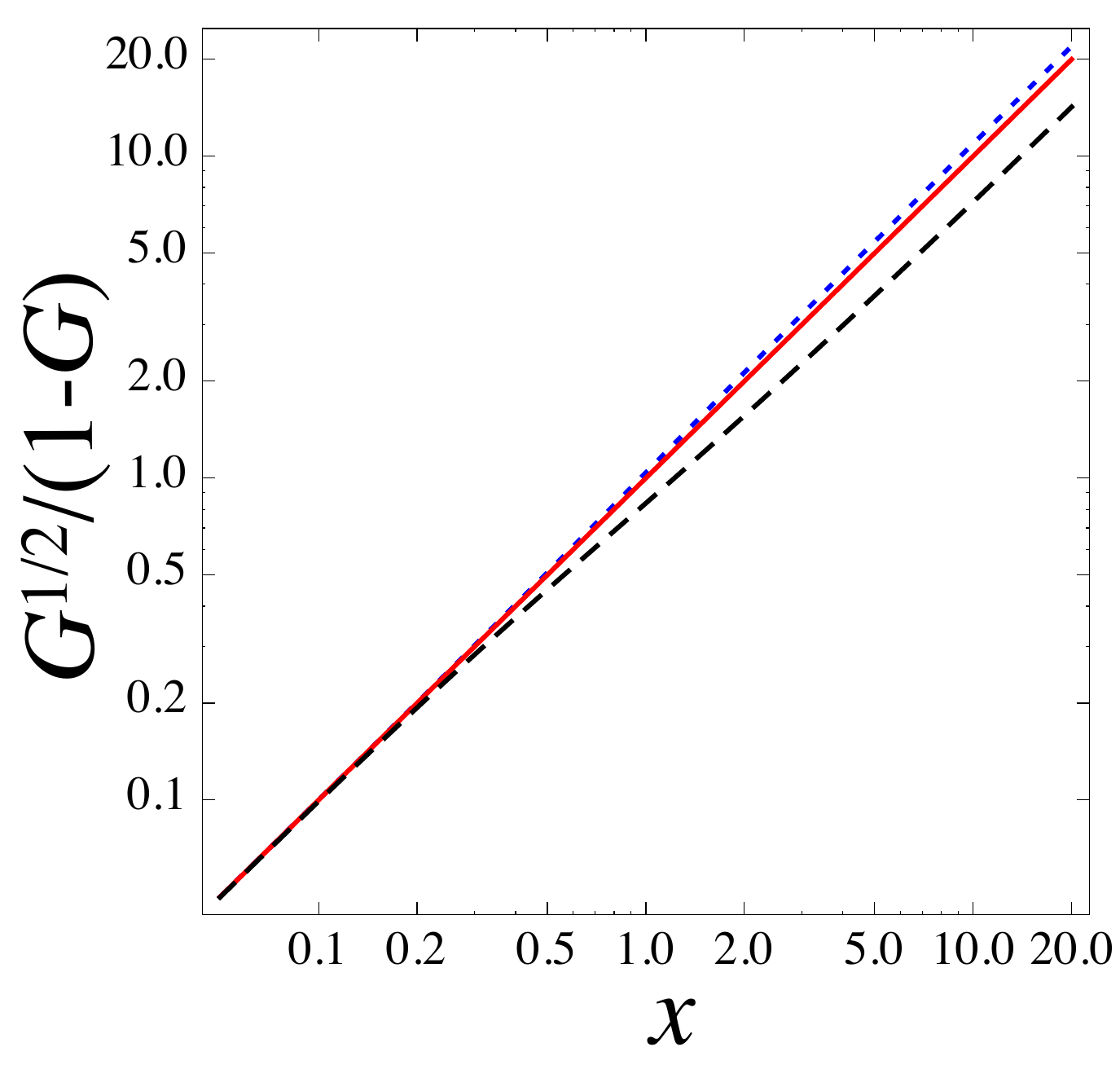}
\caption{ Comparison of the ratios  $\sqrt{G}/(1-G) $ for (i)  $G = G_{1/2}( T = x T_{1}^{\ast} )$, Eq.\ (\ref{G12V0T}), (black dashed line), (ii)   
$G = G_{1/2}( V  = x T_{2}^{\ast} )$, Eq.\ (\ref{G12VT0}),   (blue dotted line) and  the linear dependence, $\sqrt{G}/(1-G) = x $, (red solid line) expected for the expression (\ref{ourG12}). See text for additional explanations. }
\label{fig:plotK12}
\end{figure}

Secondly, in the absence of three-loop RG terms ($c_{\ast}=1$)  the ratio $G^{K}/(1-G)$, appearing in (\ref{ratio-conduc}), should be a linear function of $V$, $T$ at $K=1/2$. Plotting this ratio for the functions (\ref{G12V0T}), (\ref{G12VT0}), we compare it with the  straight line corresponding to Eq.\ (\ref{ourG12}). We confirm much better agreement with the straight line in the case of the non-linear conductance $G_{1/2}(V,T\to0)$, see Fig.\ \ref{fig:plotK12}. 

In practical terms these observations mean the following. When fitting experimental data with one universal curve for the whole range of conductances, one should use slightly different expressions for $G(V=0,T)$ and $G(V,T=0)$. The generic formula is (\ref{explicitG-c3}), where the value $c_{3}=1/4$ is appropriate for $G(V=0,T)$,   while  $c_{3}\simeq -0.07$ is better suited  for $G(V,T=0)$.
 
\subsection{Oscillatory nonlinear conductance } 

Let us also discuss the case $T=0$ and $VL$ finite.  The expression (\ref{Lambda-gen}) is reduced in this case to
\begin{equation}
\label{oscLambda}
\begin{aligned}
\Lambda &=   \ln\left( \frac{\omega _{c} e }{V} \right)  + f (2VL/v)  \,, \\ 
f(x) &= \mbox{Ci }(x) - \sin x/x  \,,\\ 
 & \simeq - \cos x/x^{2} \,,\quad x\gg1\,, \\ 
 &\simeq \gamma_{E} -1 +\ln x \,, \quad x \ll 1     \,,
\end{aligned}
\end{equation} 
 with $\gamma_{E} \simeq 0.577\ldots$. We see the appearance of oscillations in $2VL$, discussed in \cite{Dolcini2003} for the case of weak impurity. In our treatment it corresponds to $G\simeq 1$, and from 
 (\ref{explicitGLambda}) may may represent the conductance as follows 
 \begin{equation}
 G = 1 -  c_{\ast}\,  \left( {\omega _{0}}/{V } \right) ^{2 (1- K )}   \exp\left (2(1-K)   f ({2VL}/{v}) \right)
 \end{equation}
 cf.\ (\ref{GviaV}). In the limit of large $2VL/v$ we have $|f ({2VL}/{v})|\ll 1$ and  
  \begin{equation}
   \exp\left (2(1-K)   f ({2VL}/{v}) \right) \simeq 1 + 2(1-K)   f ({2VL}/{v})
 \end{equation}
 This is in  agreement with \cite{Dolcini2003} where the corresponding expression in this limit  and in our notation reads as 
 \[  
1+f^{\mbox{\small osc}}_{\mbox{\small BS}} \simeq 1 - 2(1-K) \frac{ \cos(2VL/v)}{(2VL/v)^{2}} \,.
 \]
  
\section{Conclusion \label{sec:conclusion}}

Electron transport through one-dimensional quantum wires of various types
has been studied experimentally in several recent works. In a typical set-up
stationary charge transport is measured in a two-point geometry of a system of 
one or several wires connected by a junction.  The quantum wires are
adiabatically connected to reservoirs kept at a fixed chemical potential and
temperature. These systems are described by modeling the quantum wires as
Luttinger liquids (spinless, or spinful) of fermions with linear
dispersion subject to point-like interaction and treating the reservoirs as
non-interacting. A useful picture of the transport process is to think of
individual electrons entering the interaction region (quantum wires plus
junction) from an initial reservoir and leaving as individual electrons into
the final reservoir.   If we model the reservoirs as non-interacting systems there is
no room for collective excitations such as fractional quasiparticles or
multiple quasiparticles in the final state. 

Conventionally this problem has been addressed by the bosonization method,
which takes advantage of the fact that the exact excitations of a clean
Luttinger wire are bosons, at least in the infinite wire. The problem of
including the transformation of incoming electrons into bosons has been
addressed for the clean wire, and is believed to be solved. For the case of 
semi-wires connected by a junction there is no convincing calculation of the
above transformation available. In order to avoid this difficulty we are
using a fermionic representation.

Our approach starts with determining the leading scale-dependent
contributions to the conductances in all orders of perturbation theory. We
have demonstrated in the linear response case that by summing up these terms
one arrives at a description of the critical properties near the fixed
points (i.e. the location of the FPs and the critical exponents describing
the power laws followed by the conductances). For this it is necessary to
establish the scaling property of the conductances (or else to assume its
validity, which is usually done), allowing to derive a set of
renormalization group equations out of the perturbative result. 

In the present paper we followed this approach for the case of stationary
non-equilibrium transport. We first derived a general result for the
scale-dependent terms in the conductances of an $n$-lead junction in first
order of the interaction. Then we presented the infinite order summation for
the dressed interaction. At this point we specialized our considerations to
the case of two symmetric semi-wires. We derived the corresponding
RG-equation for the conductance. In general the scaling is dependent on
three energy scales, bias voltage $V$, temperature $T$, and infrared cut-off
provided by the wire length $v/L$. Whenever one of these energy scales
dominates, the scaling variable is varying logarithmically, $\Lambda =\ln (%
\frac{v/a}{\max \{V,T,v/L\}})$. In the case $V,T \gg v/L$  we were able to
determine the form of the scaling variable describing the crossover from the
regime characterized by $V\gg T$ to $V\ll T$ , as well. 

The intermediate results presented for the general case of an $n$-lead
junction should be a good starting point for analyzing the behavior of
non-equilibrium transport through $Y$-junctions or even four-lead junctions.
Work in this direction is in progress.

\acknowledgements  
We are grateful to D. A. Bagrets, I. V. Gornyi and D. G. Polyakov for useful discussions. This work was partly
supported by the German-Israeli Foundation (GIF) and by a BMBF grant. The research of 
D.A. was supported by the Russian Scientific Foundation grant (project 14-22-00281).

\appendix

\section{Normalization of wave functions \label{sec:app:wavefunctions}}

The usual summation over the quantum states in the infinite medium is done as
an integration over the momentum $\int dk/(2\pi)$ or the summation, $\sum
_{n}$ over the quasi-momentum $k = 2\pi n/L$ in a ring geometry with a
finite length $L$. In our situation with a broken translational symmetry we
should resort to the integration over the energy, then the correct
normalization factor is given by the density of states, which is the inverse
Fermi velocity in the simplest situation, $\sum _{n} \to \int dE/v_{F}$. In
case with several Fermi velocities, $v_{j}$, in different wires we shall
keep the integration over the energy, and the normalization factor enters
the definition of the wave functions.

Thus in the formula for the retarded Green's function, 
\begin{equation}
\widehat{G}_{E}^{R}(l,y|j,x)=\int dE \frac{\phi_{E,l}(y) \phi^{\ast}_{E,j}(x)}{\omega-E+i0} ,
\label{app:defGreen}
\end{equation}
we adopt the wave functions in the $j$th wire in the form 
\begin{equation}
\phi_{E,j}(y) = e^{iE\,y/v_{j}} /\sqrt{2\pi v_{j}}
\end{equation}
and come to the formula (\ref{eq:Green}). Notice also that the integration
in (\ref{app:defGreen}) should be restricted by the electronic bandwidth $%
|E| < W =E_{F}$, which can be modeled by introducing the density of states
function, $N_{j} (E)$, with the property $N_{j} (0) = v_{j}^{-1}$. So
strictly speaking the formulas (\ref{eq:Green}) are defined at $|\omega| \ll
W$, which justifies the upper cutoff in energy in the calculation of
logarithmic corrections and the RG procedure.

\section{Keldysh structure of the triangle $T$}

The straightforward calculation shows that only a few terms in the
complicated expression for $T$ contribute to the final result. Let us sketch
here the derivation and present arguments showing the selection of the relevant terms.

To condense our writing, we use the position dependent notation, $%
G^{R}_{\omega} \to R$, and position in the product denotes the position in
the initial expression, (\ref{def:T}). So that $G^{R}_{\Omega} (z|x)
G^{L}_{\omega+\Omega}(x|y) G^{A}_{\Omega} (y|z) \leftrightarrow R L A$ etc. Up to
a numerical factor we have

\begin{equation}\begin{aligned}
T_{11} &= RAA + (K+A)(RK - RR + KA) , \\
T_{22} & = RRA+ (RK + KA + AA )(K-R) , \\
T_{21} &= RKA + KAA + RRK - RRR + AAA .  \label{eq:TviaRKA}
\end{aligned}\end{equation}
The combinations $RRR$ and $AAA$ are necessarily zero for the point $z$
outside the interacting region. We may suggest (and it is confirmed by the
direct calculation), that the contributing terms in (\ref{eq:TviaRKA}) are
those which contain two Keldysh components, $K$. In this sense, we may
keep only the terms 
\begin{equation}\begin{aligned}
T_{11} &\simeq KRK + KKA , \\
T_{22} & \simeq RKK + KAK , \quad T_{21} \simeq 0  \label{eq:TviaRKA2}
\end{aligned}\end{equation}
Note that the notation ``$\simeq$'' here also means that the combination $KK$
should be \emph{regularized} at $\Omega \to \pm \infty$ by subtracting 1
from the product of distribution factors $h_{j}(\omega+\Omega) h_{l}(\Omega)$. This
regularization is suggested by inspection of the corresponding expressions
in the direct calculation. A closer inspection shows that the combinations $%
h_{j}(\Omega) h_{l}(\Omega)$, not containing $\omega$ do not contribute to
the corrections, when multiplied by $L(\omega)$.

Thus the expressions for $T_{ij}$ can be simplified even further : 
\begin{equation}\begin{aligned}
T_{11} &\simeq KKA , \quad T_{22} \simeq RKK , \quad T_{21} \simeq 0
\end{aligned}\end{equation}
The last expression means that the corrections to the \emph{incoming}
current are absent, because $G^{A}_{\Omega} (y|z)=0$ and $G^{R}_{\Omega}
(z|x) =0$ in this case, due to the step functions in (\ref{eq:Green}).

\section{Keldysh kernel of integral equation \label{KeldyshPi}}

\begin{equation}
  \Pi ^{K}=\frac{i}{2\pi }  
 \Phi_{\omega}^{\ast} (x)
\begin{pmatrix}
\mathbf{1} F(\omega )   & \mathbf{Y}  F(\omega) \\
\mathbf{Y} F(\omega ) & \mathbf{K} (\omega) 
\end{pmatrix}   \Phi_{\omega} (y)
\end{equation}%
%
%
with 
\begin{equation*}
K_{jl}(\omega )=\int_{-\infty }^{\infty }\frac{d\Omega }{2}%
\sum_{m,n}S_{jm}^{\ast }S_{lm}S_{ln}^{\ast }S_{jn}(1-h_{m}(\Omega
)h_{n}(\Omega +\omega ))
\end{equation*}%
The latter quantity may be cast in the form
\begin{equation}\begin{aligned}
\mathbf{K} (\omega ) &=F(\omega )%
\begin{pmatrix}
1 & 0 \\ 
0 & 1%
\end{pmatrix}
+r^{2}t^{2} F_{2}(\omega,V) 
\begin{pmatrix}
1 & -1 \\ 
-1 & 1%
\end{pmatrix} \,, 
\\ 
F_{2} (\omega,V) &= F (\omega +V)+F (\omega -V)-2F (\omega ) \,.
\end{aligned}\end{equation}%
 with $F(\omega)$  defined in (\ref{def:F}). 
 Importantly, $\mathbf{K}(\omega )$ is an even function of $\omega$.

\section{Full form of the solution for $\protect\widehat{L}_{jk}^{R}(x|y) $ 
\label{sec:FullCL}}

The solution of   (\ref{eq:defC}) can be found as follows. We iterate the right-hand side of the equation once, to arrive at the diagonal kernel with components of the form 
\[   \frac {g^{2}}v \left[ \delta \left(\frac{x-z}v\right) + i \frac\omega2 (e^{i\omega(x+z)/v } + e^{i\omega |x-z|/v }) 
\right] \] 
and another component obtained from here by changing $x \to L - x$, $y \to L-y$. We pick first  the easier part of this iterated kernel, $\propto \delta(x-z)$, and arrive at the equation for $\widehat{C}_{j}(x|y)$ with more complicated inhomogeneity instead of $L_{0}$  and non-singular kernel. This latter kernel shows a jump in its derivative at $x=z$, which we use by twice differentiating    $\widehat{C}_{j}(x|y)$ with respect to $x$. We thus arrive at a second-order differential equation, similarly to what was done in \cite{Aristov2012}. The difference now is that we deal with a $2\times2$ matrix for  $\widehat{C}_{j}(x|y)$ for each wire $j$. We determine the solution to this differential equation dependent on the $x$ variable  up to terms proportional to  $e^{\pm i\omega x/(v_{j}d_{j})}$ which are multiplied by as yet unknown matrices $\hat{A}(y)$, $\hat{B}(y)$, respectively. Considering the initial Eq.\ (\ref{eq:defC}) for  $\widehat{C}_{j}(x|y)$ in the simpler cases $x=0$, $x=L$, we form a set of two coupled (matrix) equations for $\hat{A}(y)$, $\hat{B}(y)$, which is eventually solved. As a result we obtain the quantity   $\widehat{\mathbf{C}}$  diagonal in wire space, with  its diagonal elements
$\widehat{C}_{j}$ of the form 

\begin{equation}\begin{aligned}
&\widehat{C}_{j}(x|y)=\frac{2\pi v_{j}g_{j}}{d_{j}^{2}}\delta (x-y)%
\begin{bmatrix}
-g_{j}, & 1 \\ 
1, & -g_{j}%
\end{bmatrix}%
+\frac{i\pi \omega g_{j}^{2}}{d_{j}^{3}} 
   \\
&\times  e^{{i\omega |x-y|}/{v_{j}d_{j}}}
\begin{bmatrix}
d_{j}sgn(y-x)-1, & g_{j} \\ 
g_{j}, & d_{j}sgn(x-y)-1%
\end{bmatrix}
  \\
&+\frac{i\pi \omega g_{j}^{2}}{d_{j}^{4}q_{j}}\left[ 
e^{i\omega x/v_{j}d_{j}}
\hat{A}_{j}(y)+
\frac{d_{j} e^{{i\omega (2L-x)}/{v_{j}d_{j}}}}
{\sin (\frac{\omega L}{v_{j}d_{j}})}
\hat{B}_{j}(y)\right]
\end{aligned}\end{equation}%
with $d_{j}$, $q_{j}$ defined in (\ref{def:dq}) and 

\begin{equation}\begin{aligned}
\hat{A}_{j}(y) &=%
\begin{bmatrix}
g_{j}(q_{j}^{-1}-g_{j}), & g_{j}(1-g_{j}q_{j}^{-1}) \\ 
(d_{j}-1)(q_{j}^{-1}-g_{j}) , & (d_{j}-1)(1-g_{j}q_{j}^{-1})%
\end{bmatrix}
  \\
&  \times \cos (\frac{\omega y}{v_{j}d_{j}})
+id_{j}\sin (\frac{\omega y}{v_{j}d_{j}})%
\begin{bmatrix}
-q_{j}^{-1} , & 1 \\ 
-q_{j}^{-1} , & 1%
\end{bmatrix}
\\
\hat{B}_{j}(y) &=i\cos (\frac{\omega y}{v_{j}d_{j}})%
\begin{bmatrix}
\frac{(d_{j}-1)}{g_{j}} , & (1-d_{j}) \\ 
1, & -g_{j}%
\end{bmatrix}
  \\
&+d_{j}\sin (\frac{\omega y}{v_{j}d_{j}})%
\begin{bmatrix}
\frac{(d_{j}-1)}{g_{j}} , & 0 \\ 
1 , & 0%
\end{bmatrix}%
\end{aligned}\end{equation}%

We next use these expressions in Eq.\ 
(\ref{eq:LviaC}), which can be schematically represented as  
\begin{equation}\begin{aligned}
L & = C + C \ast Y \ast L  
= C + C \ast \Upsilon \ast C ,  \\    \Upsilon &= Y + Y \ast  C \ast Y + \ldots  = Y \ast (1- C \ast Y)^{-1}
\end{aligned}\end{equation}%
and obtain finally 
\begin{equation}\begin{aligned}
\label{eq:completeL}
\widehat{L}_{jk}^{R}(x|y) &=\delta _{jk}\widehat{C}_{j}(x|y)-i\omega \frac{%
2\pi g_{j}g_{k}}{d_{j}^{2}d_{k}^{2}}\Upsilon _{jk} 
\\ & \times
\begin{pmatrix}
V_{1,j}(x)V_{2,k}(y) & V_{1,j}(x)V_{1,k}(y) \\ 
V_{2,j}(x)V_{2,k}(y) & V_{2,j}(x)V_{1,k}(y)%
\end{pmatrix}%
   \\
V_{1,j}(x) &=(1-g_{j}q_{j}^{-1})\cos (\frac{\omega y}{v_{j}d_{j}}%
)+id_{j}\sin (\frac{\omega y}{v_{j}d_{j}}) \\
V_{2,j}(x) &=(q_{j}^{-1}-g_{j})\cos (\frac{\omega y}{v_{j}d_{j}}%
)-id_{j}q_{j}^{-1}\sin (\frac{\omega y}{v_{j}d_{j}})
\end{aligned}\end{equation}
with $ \Upsilon$ given in Eq.\ (\ref{integCV})

\section{Analytic properties of  $L(\omega)$.
\label{sec:analyticL}}


In contrast to our previous studies \cite{Aristov2009,Aristov2012}, 
we see now the appearance of poles in the $\omega$-plane 
of the quantities $q_{j}$, Eq.\ (\ref{def:dq}), which we
further integrate over $\omega$.  Given the arbitrariness of the $S$-matrix, reflected in $Y_{ik} = |S_{ik}|^{2}$, we check here  the absence of singularities in $L^{R}(\omega)$ in the upper semiplane of complex $\omega$. 

The poles of $q_{j}$ correspond to the solution of 
\begin{equation*}
\tan \bar \omega = - i d_{j} ,
\end{equation*}
where we introduced $\bar \omega = \omega L/{v_{j}d_{j}}$. The last equation
means that we have an infinite sequence of roots 
\begin{equation*}
\bar \omega = - i \mbox{ arctanh } d_{j} + \pi n ,\quad n = 0, \pm 1, \pm 2
, \ldots
\end{equation*}
hence the poles of $q_{j}^{-1}$ are always in the lower semiplane of complex $\omega$,
as it should be for a retarded function.

Less trivial is the question about the position of the poles of the above expression $%
(1-\mathbf{Q}^{-1} \cdot \mathbf{Y}  )^{-1}$. We consider it for a simpler situation with identical
wires, $g_{j}= g$, $d_{j} =d= \sqrt{1-g^{2}}$, $v_{j} = v$.  

The poles are defined by 
\begin{equation*} det (1-\mathbf{Q}^{-1} \cdot \mathbf{Y} )=
det \left [\mathbf{1}- \frac{g \mathbf{Y}  }{ 1 + i d \cot
\bar\omega} \right] = 0 .
\end{equation*}
Since the denominator in the last expression cannot modify the location of
poles, and $\sin \bar\omega = 0 $ is not a solution, we can rewrite 
\begin{equation*}
det \left [i d \cot \bar\omega + 1- g \mathbf{Y}  \right] = 0 .
\end{equation*}
Defining the eigenvalues of $\mathbf{Y} $ as $y_{j}$, with $j = 1,\ldots N$ (for a junction connecting $N$ wires), the poles are are defined by conditions (cf.\
above) 
$
\tan \bar \omega = - i \frac{d } {1- gy_{j} } ,
$
or 
\begin{equation*}
\bar \omega = - i \mbox{ arctanh } \frac{d } {1- gy_{j} } + \pi n ,\quad j =
1,\ldots N .
\end{equation*}

Generally we have $|y_{j}| \le 1$ for all $j$. This
is evident for $N=2$, and can be easily extended to any $N$. The proof 
of this statement is as follows. \cite{Aristov2011} Consider a set of diagonal 
$N\times N$  matrices $\lambda_{j}$ with values $1$ in the $j$th row
and $0$ otherwise. This is a set of $N$ generators of a Cartan subalgebra
of the algebra $U(N)$, normalized according to $\mbox{Tr}(\lambda_{j}%
\lambda_{k})=\delta_{jk}$. A rotation of these generators is defined as
 $\tilde\lambda_{j} = S^{\dagger} \lambda_{j} S$ where $S$ is the unitary 
 matrix. Obviously the new set  $\tilde\lambda_{j}$ remains orthonormal $\mbox{Tr}(\tilde\lambda_{j}%
\tilde\lambda_{k})=\delta_{jk}$. The operator $P$, defined by $PA = \sum_{j}
\tilde\lambda_{j} \mbox{Tr}(\tilde\lambda_{j} A)$, is a projection operator, $%
P^{2}=P$. We have $P \lambda_{k} = \sum_{j}\tilde \lambda_{j} Y_{jk}$
because  $\mathbf{Y}$ can be written as $Y_{jk} = |S_{jk}|^{2} =
\mbox{Tr}(\tilde\lambda_{j} \lambda_{k})$. Let  $\{ c_{j}\}$ be an
eigenvector of $\mathbf{Y}$, i.e.\   $\sum_{k} Y_{jk}
c_{k} = y c_{j} $. Introducing the diagonal matrix $\lambda_{\ast} = \sum_{j} c_{j} \lambda_{j}$, we 
obtain $P\lambda_{\ast} = y \tilde \lambda_{\ast}$, with $\tilde
\lambda_{\ast}$ is the rotated vector $\lambda_{\ast}$. Since $\|
P\lambda_{\ast} \| \leq \| \lambda_{\ast} \| $ and $\| \tilde \lambda_{\ast}
\| = \| \lambda_{\ast} \| $, we conclude that $|y| \le 1$.

It follows that the above ratio $\frac{d } {1- gy_{j} }$ is always
positive. 
As a result, the poles of  $(1-\mathbf{Q}^{-1} \cdot \mathbf{Y}  )^{-1}$  lie in the lower
semiplane of $\omega$.



\section{Scaling variable $\Lambda$ in the crossover regime.
\label{sec:Lambda}} 

In the limit $L\to \infty $ we have to evaluate the integral 
\begin{equation}
P(T,V) = \int _{-W}^{W}\frac{d\omega}{\omega} [F(\omega+V)-F(\omega-V) ]  \,,
\end{equation} with the upper cutoff being $W = v_{F}/a$ or the bandwidth. 

After the rescaling $\omega = 2T x$, $V=2T y$, $W=2T w$ we have 
\begin{equation*}
P(T,V) = 2T 
 \int _{-w}^{w}\frac{dx}{x} \left[ \frac{x+y}{\tanh(x+y)} - \frac{x-y}{\tanh(x-y)} \right] \,.
\end{equation*} 
We can make a shift $x \to x\pm y$  in two terms here;  the contributions from the limits $\pm w$ do not cancel upon this shift, but add a constant in the limit $ w \to \infty$. After simple calculation we get in this limit
\begin{equation}
P(T,V) = 4Ty \left( 2 + 
\int _{-w}^{w}\frac{dx \, x\coth x}{x^{2}-y^{2}} \right) \,,
\end{equation} 
where the principal value of the integral should be taken. Next we close the contour of integration in the upper semiplane of complex $x$, by adding a semicircle of radius $w$.  The contribution from this semicircle vanishes as ${\cal O}(w^{-2})$ and we reduce the remaining integral to the sum over the residues of $\coth x$ at $x=i\pi n$  as follows : 
\begin{equation} 
\int _{-w}^{w}\frac{dx \, x\coth x}{x^{2}-y^{2}}  
\to \sum_{n=1}^{w/\pi} \frac{2 n}{n^{2}+(y/\pi)^{2}}
 \,,
\end{equation} 
The last sum is easily evaluated with the final result of the form $P(T,V) = 4 V \Lambda$
with $\Lambda$ given by 
\begin{equation}
\Lambda =   \ln\left( \tfrac{W e}{2\pi T} \right)  
  - \tfrac12
\left[ \psi \left( 1 + \tfrac{iV}{2\pi T} \right) +\psi \left( 1 - \tfrac{iV}{2\pi T} \right) 
\right]    \,.
\end{equation} 
 with $e= 2.718\ldots$ and $\psi(x)$ digamma function.


\end{document}